\documentclass[a4paper]{jpconf}
\usepackage{graphicx}
\usepackage{amssymb}
\usepackage{epstopdf}
\newcommand{\beq}{\begin{equation}}
\newcommand{\eeq}{\end{equation}}

\newcommand{\Hi}{\mathcal{H}}

\newcommand{\K}{\mathcal{K}}

\newcommand{\G}{\Gamma}

\newcommand{\Hc}{{\cal H}}
\newcommand{\Aevol}{{\cal A}_{\mbox{\footnotesize{ evol}}}}

\newtheorem{definition}{Definition}

\begin{document}

\title{Conserved Quantities in Background Independent Theories}
\author{Fotini Markopoulou}

\address{Perimeter Institute for Theoretical Physics,
35 King Street North, Waterloo, Ontario N2J 2W9, Canada, and 
Department of Physics, University of Waterloo,
Waterloo, Ontario N2L~3G1, Canada}
\ead{fotini@perimeterinstitute.ca}

\begin{abstract}
We discuss the
difficulties that background independent theories based on quantum geometry
encounter in deriving general relativity as the low energy limit.
We follow a geometrogenesis scenario of a phase transition from a pre-geometric theory to a geometric phase which suggests that a first step towards the low energy limit is searching for the effective collective excitations that will characterize it.   Using the correspondence between the 
pre-geometric background independent theory and a quantum information
 processor, we are able to use the method of noiseless subsystems
to extract such coherent collective excitations.
We illustrate this in the case of locally evolving graphs. 
\end{abstract}

\section{Introduction}

The search for a quantum theory of gravity is the effort to reconcile deep
contradictions between the two fundamental theories describing nature: general
relativity and quantum theory. One such contradiction arises from the role time plays in the two theories. General relativity distinguishes itself from other theories by the fact that it is background independent. This means that its equations are invariant under the diffeomorphism group of the manifold under investigation. A canonical analysis then reveals a rather strange fact: The system is completely constrained. Instead of generating time evolution the Hamiltonian vanishes on solutions. In the quantum mechanical context this fact makes it especially hard to tackle questions of physical importance like the emergence of a classical limit. Since there is this direct connection between the background independence of general relativity and the conceptual problems of quantum gravity we want to (a) discuss the meaning of background independence and (b) show how one can make progress in the presence of background independence. 

In section \ref{sec:bi} we discuss different interpretations of background independence and order current theories accordingly. We also provide a very generic example of a background independent theory. This will allow us  to be more concrete in the following sections. Section \ref{sec:low} is devoted to the low energy limit of background independent theories. We discuss the generic problems that arise when one tries to find the classical limit of such theories. An intriguing possibility is discussed in section \ref{sec:genesis}. In a background independent theory it is possible that the means to solve the low energy problem are themselves emergent. This leads to the concept of geometrogenesis, i.e. the genesis of spacetime itself. 

We suggest that, starting from a background independent theory, we may infer a spacetime through the properties  of emergent coherent degrees of freedom. Quantum information theory provides a tool that is very well suited to finding coherent degrees of freedom for background independent theories. This is the method of  noiseless subsystems. In section \ref{sec:noise} we introduce noiseless subsystems rigorously and discuss their use.  Finally in section \ref{sec:disc} we discuss how a different understanding of background independence may provide a completely different way of arriving at a quantum theory of gravity.

%


\section{Background Independent quantum theories of gravity}\label{sec:bi}

Background independence (BI) is thought to be an important part of a quantum theory of gravity since it is an important part of the classical theory \cite{Ish,Sta,Smo,Rov}.  Background independence in general relativity is the fact that physical quantities are invariant under spacetime diffeomorphisms.  There is no definite agreement on the form that BI takes in quantum gravity.  Stachel \cite{Sta} gives  the most concise statement of background independence:  ``In a background independent theory there is no kinematics independent of dynamics''.

In the present article, we shall need to discuss specific aspects of background independence and to aid clarity we give the following two definitions that we shall use:

\begin{definition}
Background independence I (BI-I):
A theory is background independent if its basic quantities and concepts do not presuppose the existence of a background spacetime metric.  
\end{definition}

All well-developed background independent approaches to quantum gravity such as Loop quantum gravity \cite{LQG}, causal sets \cite{CauSet}, spin foams \cite{SF}, causal dynamical triangulations \cite{DT}, dynamical triangulations \cite{CDT} or quantum Regge calculus \cite{QRC} implement background independence as a special case of the above by quantum analogy to the classical theory:

\begin{definition}
Background independence II (BI-II):  A background independent theory of quantum geometry is characterized
by a) quantum geometric microscopic degrees of freedom or a regularization of the microscopic geometry and b) a quantum sum-over-histories of the allowed microscopic causal histories (or equivalent histories in the Riemannian approaches).  
\end{definition}

Recently, new approaches to quantum gravity have been proposed that satisfy BI-I but not BI-II:  the computational universe \cite{Llo}, internal relativity \cite{Dre} and quantum causal histories \cite{TGQ}.  

Presenting any one of these theories is  beyond the scope of this paper and reviews can be found in the literature. 
Our focus is the connection between a BI theory and our low-energy world, problems encountered in attempting to establish such a connection (the low-energy problem) and recent progress in extracting conserved quantities from such theories.  For this purpose it will suffice to look at an example of a basic background independent theory which is very simple but sufficient for the analysis in this article.

\subsection{Example of a background independent theory}
\label{thetheory}

Possibly the most common objects that appear in background independent theories are graphs.  Graph-based, instead of metric-based, theories are attractive implementations of the relational content of diffeomorphism invariance:  it is the connectivity of the graph (relations between the constituents of the universe) that matter,  not their distances or metric attributes.
We'll use a very simple  graph-based system to illustrate the low energy limit problem for background independent systems and show how to find conserved quantities in such theories.  

We start with a graph $\G$ of $n=1,\ldots,N$ nodes, each with three edges attached to it, embedded in a topological 3-dimensional space $\Sigma$ (no metric on $\Sigma$).  A map from $\G$ to a quantum system can be made by associating a finite-dimensional state space $\Hc_n$ to each minimal subgraph of $\G$, namely, the constitutent open graphs containing one node and three open edges:
\beq
	\Hc_n=\begin{array}{c}\mbox{\includegraphics{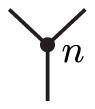}}\end{array}.
\eeq
The state space of the graph $\G$ is the tensor product over all the constituents\footnote{If there are labels on the edges of the graph, the space (\ref{eq:HG}) is   $\Hc_\Gamma=\sum_{e\in\Gamma}
\bigotimes_{n\in\G}\Hc_n$ where there sum is over labels on the edges $e$ connecting the nodes. },
\beq
\Hc_\Gamma=\bigotimes_{n\in\G}\Hc_n,
\label{eq:HG}
\eeq
and the state space of the theory is 
\beq
\Hc=\bigoplus_{\G_i}\Hc_{\G_i},
\label{eq:H}
\eeq
where the sum is over all topologically distinct embeddings of all such surfaces in
$\Sigma$ with the natural inner product $\langle\G_i|\G_{i'}\rangle=\delta_{\G_i\G_{i'}}$.

\begin{figure}
\begin{center}
\begin{equation}
\begin{array}{rl}
A_1&=\begin{array}{c}\includegraphics{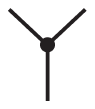}\end{array}
\longrightarrow
\begin{array}{c}\includegraphics{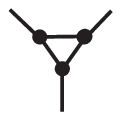}\end{array}
\\
A_2&=\begin{array}{c}\includegraphics{Hntriangle.eps}\end{array}
\longrightarrow
\begin{array}{c}\includegraphics{Hn1.eps}\end{array}
\\
A_3&=\begin{array}{c}\includegraphics{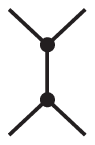}\end{array}
\longrightarrow
\begin{array}{c}\includegraphics{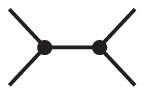}\end{array}
\end{array}
\end{equation}
\end{center}
\caption{
The three generators of evolution on the graph space $\Hc$. They are called expansion,
contraction and exchange moves.}
\label{fig:moves}
\end{figure}

Local dynamics on ${\cal H}$ can be defined by excising subgraphs of $\Gamma$ and replacing them with new ones with the same boundary.  The generators of such dynamics are given graphically in Fig.\ref{fig:moves} \cite{Mar97,MarSmo97}. 
 Given a  graph $\Gamma$, application of $A_i$ results in 
 \beq
\hat{A}_i |\Gamma \rangle = \sum_{\alpha} |\Gamma^\prime_{\alpha i}\rangle
\eeq
 where $\Gamma^\prime_{\alpha i}$ are all the graphs obtained
 from $\Gamma$ by an application of one move of type $i$  ($i=1,2,3$).  
Together with the identity ${\bf 1}$, these moves generate the {\em evolution algebra}
\beq
\Aevol=\left\{{\bf 1},A_i\right\}, \qquad i=1,2,3
\eeq
 on ${\cal H}$ . 

An example of such a system is spin networks \cite{Pen,RovSmo,MajSmo} where the graph edges are labeled by representations of $SU(2)$ or $SU_q(2)$ and $\Hc_n$ is the space of intertwiners for the representations on the edges of node $n$ (a one-dimensional space for $SU(2)$ spin networks with three edges on each node but non-trivial for more edges or $SU_q(2)$ graphs).  Spin networks evolving under $\Aevol$ appear in spin foam models \cite{SF}.

We can summarize the features of such systems:
\begin{itemize}
\item
Graph-based states (no background metric in $\Hc$).
\item
Locally finite (implemented by finite-dimensional spaces $\Hc_n$). 
\item
Evolution local with respect to the graph, as illustrated in Fig.\ref{fig:graphs}.  As has been analyzed in \cite{HawMarSah, LivTer}, the boundary preserving operations in $\Aevol$ are unitary but more detailed {\em completely positive} evolution operators that do not preserve boundaries are also sufficient ($\Phi$ in Fig.\ \ref{fig:graphs}).  Completely positive maps are commonly used to describe evolution of open quantum systems and are defined as follows \cite{NieChu}.
\begin{figure}
\begin{center}
\includegraphics[height=3cm]{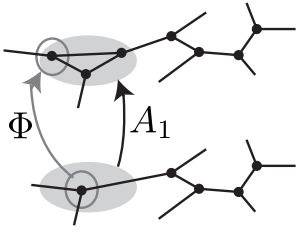}
\end{center}
\caption{Local evolution of graphs.  The boundary-preserving operator $A_1$ is unitary while $\Phi$ is completely positive \cite{HawMarSah}.}
\label{fig:graphs}
\end{figure}

Let $\Hi_S$ be the state space of a quantum system in contact with
an environment $\Hi_E$ (here $\Hc_S$ is the subgraph space and $\Hc_E$ the space of the rest of the graph). 
The standard characterization of evolution
in open quantum systems starts with an initial state in the system
space that, together with the state of the environment, undergoes
a unitary evolution determined by a Hamiltonian on the composite
Hilbert space $\Hi = \Hi_S \otimes \Hi_E$, and this is followed by
tracing out the environment to obtain the final state of the
system. 

The associated evolution map
$\Phi:{\cal A}(\Hi_S)\rightarrow{\cal A}(\Hi_S)$ between the corresponding matrix algebras of operators on the respective Hilbert spaces  is necessarily completely
positive (see below) and trace preserving. More generally, the map
can have different domain and range Hilbert spaces. Hence the
operational definition of quantum
evolution $\Phi$ from a Hilbert space $\Hi_1$ to
$\Hi_2$ is:
\begin{definition} Completely positive (CP) operators.
A {\em completely positive} map  $\Phi$ is a linear map 
$\Phi: {\cal A}(\Hi_1)\rightarrow{\cal A}(\Hi_2)$ such that the maps
\beq
 id_k \otimes \Phi :  M_k \otimes {\cal A}(\Hi_1) \rightarrow
 M_k \otimes {\cal A}(\Hi_2)
\eeq
are positive for all $k\geq 1$. 
\end{definition}
Here we have written $M_k$ for
the algebra ${\cal A}(\mathbb{C}^k)$ represented as the $k\times k$
matrices with respect to a given orthonormal basis. (The CP
condition is independent of the basis that is used.)

\item
$\Hc$ contains information about the topology of the embedding of $\G$ in $\Sigma$.  
\item
A sequence of moves in $\Aevol$ results in a history.  At this point there is an important choice to  be made:  

1.  The dynamics of the theory is given by a superposition of all allowed such histories.  This is the case in all BI-II type theories.  In essence, each history is one spacetime geometry and the quantum theory involves the superposition of all of these.  

2.  There is only one history.  This is the case in the computational universe, internal relativity and quantum causal histories.  These claim to be BI-I theories where the history is a {\em pre-geometric} quantity.  The meaning of these is briefly  discussed further in the conclusion.

\end{itemize}

\section{The low energy limit of background independent theories}\label{sec:low}

A background independent approach to quantum gravity will be a successful quantum theory of gravity if  a) it is well-defined, b) general relativity (and the quantum field theory of the appropriate matter content) emerges as as the low-energy limit of the theory and c) it makes predictions on the kind and magnitude of departure from the classical theory.    

Without going in detail into specific issues that arise in each of the BI approaches to quantum gravity, one can get an idea of the problems that one encounters in the quest for the low energy limit of background independent theories, especially issues specific to BI systems by comparing our example to a condensed matter system.  
The graph $\G$  plays the role of the lattice, while the ${\Hc}_n$'s are the microscopic quantum degrees of freedom.  The low energy problem is analogous to describing the macroscopic behaviour emergent from a many-body system in condensed matter physics.  Building on that analogy, there has been work, for example, on the application of renormalization group methods to such BI systems \cite{Mar00,RG}.  

There are, of course, technical obstacles such as the irregular nature of the lattices, the often complicated calculations involving the microscopic variables (usually group representations) and the lack of experimental controls,  readily available in standard condensed matter systems.  But there are also problems specific to BI systems:

\begin{itemize}

\item
{\em Dynamics}.
The low energy behavior of a physical  system depends on its dynamics.  

Causal Dynamical Triangulations (CDT)  is a clear demonstration of this basic fact of physical systems in quantum gravity.  Both CDT and Euclidean Dynamical Triangulations  (DT) \cite{DT} start with building blocks of the same dimensionality, four-simplices.  They differ in the dynamics.  In the continuum limit, CDT finds Hausdorff and heat dimensions near 3+1, while the Euclidean theory ends up either with effective dimension of 2 or infinite\footnote{A concise explanation of this is given by R.\ Loll \cite{Lolweb}: ``How can one obtain an effective geometry that is not four-dimensional by superposing virtual spacetime geometries that individually are of dimension four? This nonperturbative quantum phenomenon, completely at odds with our classical intuition of spacetime as a fixed inert background structure, beautifully illustrates that no aspect of such a dynamically generated geometry can be taken for granted, not even its dimension. Since all local curvature degrees of freedom of the geometry undergo large quantum fluctuations, a four-dimensional geometry can either crumple up to generate a geometry of an effectively higher dimension (like crumpling up a two-dimensional sheet of paper into a three-dimensional ball), or curl up to give a geometry of an effectively lower dimension (like rolling up a piece of paper into a thin tube, which will appear effectively one-dimensional at a scale much larger than the circumference of the tube). This is exactly what happens in nonperturbative Euclidean quantum gravity. It produces ground states of geometry that are either maximally crumpled with an infinite(!) effective dimension or polymerized into thin and branched threads, with an effective dimension of two, neither of them promising candidates for the ultimate vacuum.'' 
}.

Dynamics is notoriously difficult to implement in most background independent approaches, which makes it tempting to draw conclusions about the physical content of a theory before we have taken dynamics into account.  For example, spin foam models often relate the valence of the nodes in the spin foam 2-complex to the dimensionality of the system and much of the analysis of specific models involves analyzing the properties of a single building block  without considering the entire path-integral.  This is analogous to considering a spin system in condensed matter physics and inferring properties of its continuum limit by looking at the spins,  independently of the hamiltonian.  The Ising model in 2 dimensions \cite{Isi} and string networks \cite{Wen} have precisely the same building blocks and kinematics, square lattices of spins, but different dynamics.  The resulting effective theories could not be more different.   In the field of quantum gravity itself, the example of CDT vs DT shows us how little trust we should put in properties of the microscopic constituents surviving to the low-energy theory.  

We must conclude that any method we may use to analyze the low-energy properties of a theory needs to take the dynamics into account.

\item
{\em Observables}.  
Using the analogy between the graphs $\G$ of our theory and a condensed matter system, we may consider applying condensed matter methods to the graphs, such as a real space renormalization (coarse-graining the graph).  However, careful inspection of the real space renormalization method in ordinary systems shows that 
implicit in the method is the fact that, coarse-graining the lattice spacing coarse-grains the observables.  
In BI systems, the best we can do is relational observables and there is 
 no direct relationship between BI observables and the lattice or the history.  Hence,  the physical meaning of coarse-graining a graph is unclear. 
 
 In  theories of regularized geometries, such as CDT, there is a somewhat different issue.  The continuum limit observables that have been calculated so far are averaged ones, such as the Hausdorff or heat dimensions.  One still needs to find localized observables in order to compare the predictions of the theory to our world.  

\item
{\em (Lack of) symmetries}.
We should clarify that when we use the term {\em low-energy} it is only by analogy to ordinary physics and both {\em energy} and {\em low} are ill-defined.  The definition of energy needs a timelike Killing vector field, clearly not a feature of a BI theory.  A notion of scale is necessary to compare {\em low} to {\em high}.  Outside CDT, it is not clear how scale enters BI systems.

\item
{\em Time}.
This is another facet of the dynamics/observables problem.  BI-II theories, with the exception of CDT, are timeless leading to the three problems outlines above.  There may be a way out for BI-I theories, as we discuss in the concluding section and elsewhere \cite{KonMar,TGQ}.

\end{itemize}

Note that all of the above issues that we blocked under ``dynamics'', ``observables'', ``symmetries'' and ``time'' are really different aspects of the question of dynamics in background independent theories.

\section{The geometrogenesis picture}\label{sec:genesis}

Let us consider a simple scenario of what we may expect to happen in a BI theory with a good low energy limit.   It is a factor of about twenty orders of magnitude from the physics of the Planck scale  described by the microscopic theory to the standard subatomic physics.  By analogy with all other physical systems we know, it is reasonable to expect that physics at the two scales decouples to a good approximation.  We can expect at least one phase transition interpolating between the microscopic BI phase and the familiar one in which we see dynamical geometry.  We shall use the word {\em geometrogenesis} for this phase transition.

Admittedly, we only have guesses as to the microscopic theory and very limited access to experiment.  Additionally, phase transitions are not very well understood even in ordinary lab systems, let alone phase transitions of background independent systems. In spite of these issues, we find that the geometrogenesis picture suggests a first step towards the low energy physics that we can take.  

A typical feature of a phase transition is that the degrees of freedom that characterize each of the two phases are distinct (e.g.\  spins vs spin waves in a spin chain or atoms vs phonons in solid state systems), with the emergent degrees of freedom being collective excitations of the microscopic ones.  In our example, the vector spaces on graphs contain the microscopic degrees of freedom and operators in $\Aevol$ is the microscopic dynamics.  Is there a way to look for collective excitations of these that are long-range and coherent so that they play a role in the low-energy phase? 

We find that this is possible, at least in the idealized case of conserved (rather than long-range) quantities in a background independent system such as our example.   The method we shall use, {\em noiseless susbsystems}, is borrowed from quantum information theory, thanks to a straightforward mapping that exists between locally finite BI theories and quantum information processing systems.   We present this mapping as well as the method used in the next section.  

We are then suggesting a new path to the effective theory of a background independent system \cite{KriMar,TGQ,KonMar,KonMarSmo}.  The basic strategy is to begin by identifying effective coherent degrees of freedom and use these and their interactions to characterize the effective theory.  If they behave as if they are in a spacetime, we have a spacetime.

\section{The quantum information theoretic perspective}\label{sec:noise}

\begin{figure}
\begin{center}
\includegraphics[height=4cm]{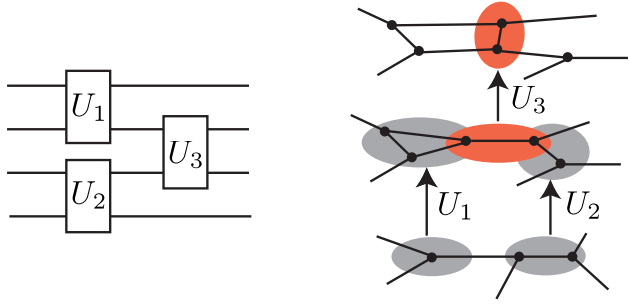}
\end{center}
\caption{A quantum information processing circuit (L) and the locally evolving graphs (R).}
\label{fig:graphs2}
\end{figure}

What does a quantum circuit and our example theory have in common?  They are the same mathematical structure, tensor categories of finite-dimensional vector spaces with arrows that are unitary or CP operators (see Figs.\ \ref{fig:graphs} and \ref{fig:graphs2}) \cite{Mar99a,Mar99b,HawMarSah,LivTer}.  This is also the case with a variety of other theories,  from topological quantum field theory to string networks in condensed matter physics \cite{Wen}.  This is simply the mathematics of finite dimensional quantum systems.  What is interesting for us is that this mathematics contains no reference to any background spacetime that the quantum systems may live in and hence it is an example of  BI-I.

In \cite{MarPou,KriMar}, we found that the field of quantum information theory has a notion of coherent excitation which, unlike the more common ones in quantum field theory and condensed matter physics,  makes no reference to a background geometry and can be used on a BI system.  
This is the notion of a {\em noiseless subsystem} (NS) in quantum error correction, a subsystem protected from the noise, usually thanks to symmetries of the noise \cite{NS}.  Our observation is that passive error correction is analogous to problems concerned with the emergence and
stability of  persistent quantum states in condensed matter physics.     
In a quantum gravity context, the role of noise is simply the fundamental evolution and the existence of a noiseless subsystem 
means a coherent excitation protected from the microscopic
Planckian  evolution, and thus relevant for the effective theory. 

\begin{definition}
Let $\Phi$ be a quantum channel on $\Hc$ and suppose that $\Hc$ decomposes as 
$\Hc =(\Hc^A\otimes\Hc^B)\oplus\K$, where $A$ and $B$ are subsystems and
$\K =(\Hc^A\otimes\Hc^B)^\perp$. We say that $B$ is {\em noiseless}
for $\Phi$ if
\begin{eqnarray}\label{ns}
\forall\sigma^A\ \forall\sigma^B,\ \exists \tau^A\ :\
\Phi(\sigma^A\otimes\sigma^B) = \tau^A\otimes \sigma^B.
\end{eqnarray}
\label{eq:NS}
\end{definition}
Here we have written $\sigma^A$ (resp.\ $\sigma^B$) for operators
on $\Hc^A$ (resp.\ $\Hc^B$), and we regard $\sigma = \sigma^A\otimes
\sigma^B$ as an operator that acts on $\Hc$ by defining it to be
zero on $\K$.  Note that, given $\Hi$ and $\Phi$, it is a non-trivial problem to find a decomposition that exhibits a NS.  Much of the relevant literature in quantum information theory is concerned with algorithmic searches for a NS given $\Hi$ and $\Phi$.  

The noiseless subsystem method (also called decoherence-free
subspaces and subsystems) is the fundamental passive technique for
error correction in quantum computing.  In this setting, the operators $\Phi = \{E_a\}$ in the
operator-sum representation for a channel are
called the {error} or {noise} operators associated with
$\Phi$. It is precisely the effects of such operators that must be
mitigated for in the context of quantum error correction \cite{NS}.
The basic idea in this setting is to (when
possible) encode initial states in sectors that will remain immune
to the deleterious effects of the errors $\Phi = \{E_a\}$
associated with a given channel.

The term ``noiseless'' may be confusing in the present context:  it is not necessary that there is a noise in the usual sense of a given split into system and environment.  As is clear from the definition that follows, simple evolution of a dynamical system is all that is needed, the noiseless subsystem is what evolves coherently under that evolution.

In conclusion, our basic idea is that, if there is an effective theory, it is characterized by effective degrees of freedom which remain largely coherent, i.e., protected from the Planckian evolution and hence relevant for the low energy limit.  It is easiest to search for such excitations in an idealized setup in which they are left completely invariant under the evolution generated by $\Aevol$.  
NS is a method to identify these.  Next, we illustrate the method by applying it to our example theory.

\subsection{Noiseless subsystems in our example theory}

Are there any non-trivial noiseless subsystems in $\Hi$?   There are, and are revealed when we rewrite $\Hi_S$ in eq.\ (\ref{eq:H}) as
\beq
\Hi_S=\Hi_S^{n'}\otimes\Hi_S^B,
\label{eq:Hnew}
\eeq
where $\Hi_S^{n'}:=\bigotimes_{n'\in S}\Hi^{n'}$ contains all {\em unbraided} single node subgraphs in $S$ (the prime on $n$ serves to denote unbraided) and $\Hi_S^b:=\bigotimes_{b\in S}\Hi_b$ are state spaces associated to braidings of the edges connecting the nodes.  For the present purposes, we do not need  to be explicit about the different kinds of braids that appear in $\Hi_S^b$.  

The difference between the decomposition (\ref{eq:H}) and the new one (\ref{eq:Hnew}) is best illustrated with an example (details can be found in \cite{BilMarSmo}).  Given the state
\beq
    \begin{array}{c}\mbox{\includegraphics{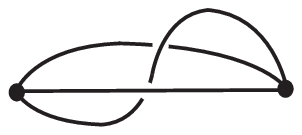}}\end{array}
\eeq
eq.\ (\ref{eq:H}) decomposes it as 
\beq
    \begin{array}{c}\mbox{\includegraphics{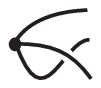}}\end{array}\otimes
     \begin{array}{c}\mbox{\includegraphics{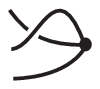}}\end{array}
\eeq
while (\ref{eq:Hnew}) decomposes it to 
\beq
       \begin{array}{c}\mbox{\includegraphics{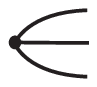}}\end{array}\otimes
    \begin{array}{c}\mbox{\includegraphics{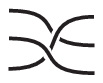}}\end{array}\otimes
     \begin{array}{c}\mbox{\includegraphics{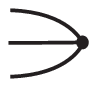}}\end{array}.
\eeq

With the new decomposition, one can check that operators in $\Aevol$ can only affect the $\Hi_S^{n'}$ and that $\Hi_S^b$ is {\em noiseless under} $\Aevol$.  This can be checked explicitly by showing that the actions of braiding of the edges of the graph and the evolution moves commute.  

We have shown that braiding of graph edges are unaffected by the usual evolution moves.  Any physical information contained in the braids will propagate coherently under $\Aevol$.  These are effective coherent degrees of freedom.  The physical interpretation of the braids is beyond the scope of this paper (see \cite{BilMarSmo}) for an interpretation of the braids as quantum numbers of the standard model).  

Note that this example may appear simple but the fact that the widely used system of locally evolving graphs exhibits broken ergodicity \cite{Pal}  (namely $\Hc$ splits into sectors, characterized by their braiding content, and $\Aevol$ cannot take us between sectors) went unoticed prior to the introduction of the NS method.

\section{Discussion and Conclusions}\label{sec:disc}

We started by defining the different kinds of background independent theories that we study in this article.  We make a distinction between theories that are background independent in the straightforward sense that no fixed geometric quantities appear in their basic formulation (BI-I) and implementations of background independence via superpositions of quantum geometry (BI-II).  We noted that BI-II theories are naturally sum-over-histories theories while BI-I may be given in terms of a single pre-geometric history.  
We gave a simple example of a BI-I theory, evolving graphs of locally finite quantum systems.  

We then discussed issues that arise when we attempt to extract the low-energy physics of BI systems.  In particular, we find that dynamics in its various guises (observables, symmetries, time, etc) both gives rise to difficult problems {\em and } it is essential in determining the properties of the low energy theory. 

We sketched out geometrogenesis, a phase transition scenaio whose microscopic phase is a BI system and the macroscopic phase is general relativity plus matter.  This led us to focus on collective coherent excitations of the BI theory as a first step towards being able to describe such a transition.  

Finally, we took advantage of the similarities between locally finite BI systems and a quantum computation to import a method from quantum information, noiseless subsystems, to our problem.  Noiseless subsystems provide a way to identify collective conserved quantities in a BI theory which can be thought of as a special case of long-range propagating degrees of freedom.  We illustrated this in our graph-based system where we found that the commonly used expansion, contraction and exchange evolution operators conserve braiding information.  

There are a number of questions we have left open.  We will comment on two:

1.  Scale in a background independent theory. 
Our conserved quantities can be thought of as a special case of long-range propagating degrees of freedom, where the lifetime of the propagating ones is infinite.  Noiseless subsystems can only deal with this case because it only looks at the symmetries of the microscopic dynamics (see \cite{KriMar}).  This is not a realistic picture for what we are looking for, i.e.\ the degrees of freedom in the geometric phase.  Presumably, what we need is to weaken the notion of a NS to ``approximately conserved'' so that it becomes long-range rather than infinite.  Long-range, however, is a comparative property and to express it we need a way to introduce scale into our system.  It is unclear at this point whether it is possible to introduce a scale in a BI theory without encountering the problems listed in section 3.  

2.  BI-I vs BI-II.  
The geometrogenesis picture leads one to reconsider the role of microscopic quantum geometric degrees of freedom traditionally present in BI theories (BI-II).  It appears unnatural to encounter copies of the geometry characteristic of the macroscopic phase already in the microscopic phase, as is the case, for example, when using quantum tetrahedra in a spin foam.  
Instead, one can start with a BI-I type theory with no notion of geometry present and look for the effective coherent degrees of freedom along the lines we described.  If these and their interactions behave as if they are in a spacetime with gravity  then we have a geometric and gravitational macroscopic phase.  

The advantage of this viewpoint is mainly that it allows for ordinary quantum dynamics, instead of a Wheeler-deWitt evolution, potentially providing a way out of the issues listed in section 3.  The obvious two difficulties are: a)  Time.  Does this ordinary dynamics amount to a background time?  Recent work \cite{Dre, Llo, KonMar} indicates that the answer is not clear.  In particular, \cite{Dre, Llo} argue that the very absence of fundamental geometry and the accompanying separation between geometry and matter will lead to a dynamical spacetime in the low energy.  b) How can we get geometry and gravity out if we do not put them in?  In essence, a BI-I theory needs to {\em explain}  the Einstein equations, not an ambition for BI-II theories which aim to give a consistent quantization of the Einstein equations.  Suggestions on how this may be possible can be found in \cite{Dre,TGQ}.

\section*{Acknowledgements}

The material in this article is extracted from joint work with Olaf Dreyer, Tomasz Konopka, David Kribs, David Poulin and Lee Smolin.  I am also grateful to Florian Girelli, Etera Livine, Dan Gottesman, Danny Terno, Paolo Zanardi and
my colleagues at PI for many fruitful discussions. Research at
PI is supported in part by the Government of Canada through NSERC and
by the Province of Ontario through MEDT. Parts of this work were supported by a grant from FQXi.

\end{document}